# Flickering of buoyant diffusion flames: a vortex dynamics revisit

Xi Xia and Peng Zhang[†]

*Department of Mechanical Engineering*

*The Hong Kong Polytechnic University, Hung Hom, Hong Kong*

**Abstract:** Previous studies have found that the flickering of buoyant diffusion flames is associated with the periodic shedding of a toroidal vortex, which is formed under gravity-induced shearing at the flames. Moreover, numerous experimental investigations have reported the scaling relation, $f \propto D^{-1/2}$, where $f$ is the flickering frequency and $D$ is the diameter of the fuel inlet. However, the connection between the toroidal vortex mechanism and the scaling relation has not been clearly understood. The current study theoretically revisits this problem from the perspective of vortex dynamics. The theory incorporates a recent finding in the vortex dynamics community: the detachment of a continuously growing vortex ring is inevitable and can be dictated by a universal constant that is essentially a non-dimensional circulation of the vortex. By calculating the total circulation of the toroidal vortex and applying the vortex ring detachment criterion, we mathematically established the connection between the lifetime of a toroidal vortex and the flickering of a buoyant diffusion flame, resulting in a general flickering frequency formulation, which is validated by comparing with existing experimental data of pool and jet diffusion flames.

**Keywords:** Buoyant diffusion flames; Flickering frequency; Toroidal vortex; Vortex-sheet formation; Vortex detachment.

---

[†]Corresponding author: pengzhang.zhang@polyu.edu.hk

# I. Introduction

Diffusion flames are constantly involved in domestic and industrial applications that have been shaping the human civilization, including energy production, propulsion, and fire protection[1, 2]. Diffusion flame characteristics are dominated by the convection and diffusion of species in the flow field where the flame is embedded, and are less affected by the chemical reactions that usually happen within a short period and inside a thin layer. In practice, diffusion flames tend to become unstable under the effect of buoyancy [3-12], which could significantly impact combustion performance and fire safety. A prominent phenomenon related to the stability of a buoyant diffusion flame is the flame flickering [13] or puffing, the oscillation frequency $f$ of which has been a constant subject of interest for several decades.

Early researchers [14-19] experimentally found that $f$ is related to the fuel source dimension $D$ as $f \propto D^{-1/2}$ with a proportionality factor of around 1.5 given by Trefethen and Panton [20] and Cetegen and Ahmed [21], although slightly different values of the factor have been reported by other studies [8, 9, 15, 18]. Based on dimension analysis, Byram and Nelson [15] provided the first theoretical derivation of the scaling law in the form of $f \propto (g/D)^{1/2}$, with $g$ being the gravitational acceleration. This scaling law was shown to be equivalent to the Strouhal-Froude number correlation, $St \propto Fr^{-1/2}$, by Hamins et al. [22] and Malalasekera et al. [8]. The same scaling law was also obtained by Bejan [23] with the use of buckling theory of inviscid streams. Cetegen and Ahmed [21] derived the scaling law by assuming the pressure to reach a hydrostatic equilibrium across the flame sheet; they further extended it to account for a general buoyant jet diffusion flame by including the Richardson number, which measures the gravitational effect relative to the initial fuel momentum.

The previous investigations point to a striking feature of buoyant diffusion flames that the oscillation frequency is flow-dominated, meaning that its essence is less affected by the chemical aspects of the flame. To understand the fundamental mechanism causing the flame flickering, Buckmaster and Peters [24] performed a linear stability analysis to an infinite candle model and attributed the flickering to a Kelvin-Helmholtz type of instability. Chen and Roquemore [3, 25, 26] together with their collaborators presented the visualizations of the flow structures within jet diffusion flames and identified two different types of vortices — the small vortices inside of the luminous flame developed due to the instability of the jet and the large toroidal vortices outside of the flame sheet caused by the buoyancy-induced Kelvin-Helmholtz instability. Moreover, Chen et al. [3] attributed the flame oscillation to the dynamics of the toroidal vortices, which were observed to result in the detachment of the flame puff (flame pinch-off [27, 28]) as well as "pairing" and "merging" of the flame bulge, as illustrated in Fig. 1. This finding was also confirmed by the numerical simulation of Katta and Roquemore [29]. Cetegen and Ahmed [21] further substantiated this view by predicting the oscillation frequency based on a convective time scale associated with the toroidal vortex. Subsequently, numerous experimental and numerical investigations have been conducted to study the dynamics of the toroidal vortex and its interaction with flame [10, 30-33]. It is quite clear from these studies that flickering is primarily caused by buoyant flow instability and the toroidal vortex, at least at sufficiently small Reynolds numbers, while there has been a debate [10, 34-36] on whether the nature of the instability is absolute or convective.

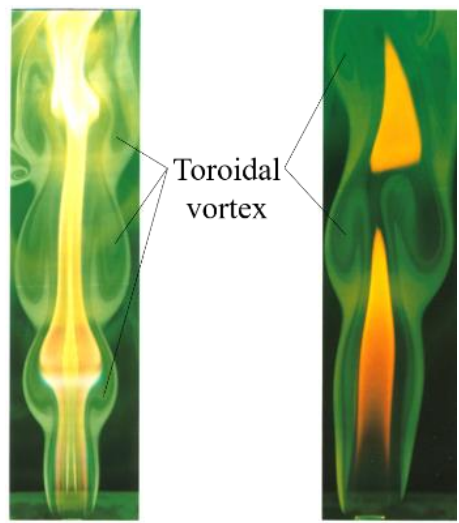

Fig.1 Flow visualizations of laminar jet diffusion flames [3] show the synchronization between the flame structures and the toroidal vortices.

To this day, the jigsaw puzzle of the flame flickering has almost been completed, left with only a few unsolved pieces, one of which is the exact connection between the toroidal vortex and the flame oscillation frequency. Although Cetegen and Ahmed [21] have showed that the oscillation frequency is consistent with the convective time scale of the toroidal vortex, the detailed understanding and physical picture of the mechanism are still not clear. Particularly, the germane and critical problem is to mathematically elucidate how the complex vortex dynamics, including the generation of vorticity, the roll-up of the toroidal vortex, and the vortex detachment, can be integrated into a single model that quantitatively determines the flame oscillation. To this end, a primary task of the current study is to establish a mathematical model based on toroidal vortex to predict the frequency of a buoyant diffusion flame. The theories developed in this work will be validated against the frequency data of various pool and jet diffusion flames reported by previous experimental and numerical studies.

**II. Mathematical Model Based on Toroidal Vortex**

As stated above, the essence of the flame flickering or puffing lies in the dynamics of the toroidal vortex. Naturally, the first question to ask is how the toroidal vortex forms. In vortex dynamics, the formation and evolution of toroidal vortex, formally known as the "vortex ring", have been studied extensively [37-43]. Physically, the appearance of a vortex ring may be considered as the outcome of the growing and rolling-up of a cylindrical-shaped vortex sheet. A well-known example is the starting vortex jet [40, 43, 44], where the vortex sheet is continuously supplied by pushing a fluid slug out of a circular jet nozzle to form a cylindrical shear layer. From this perspective, the toroidal vortex of a diffusion flame should not be fundamentally different and its formation must involve a growing vortex sheet. In fact, previous experimental [3] and numerical [10] studies have shown that a strong vortex sheet coinciding with the flame sheet is responsible for the formation of the toroidal vortex.

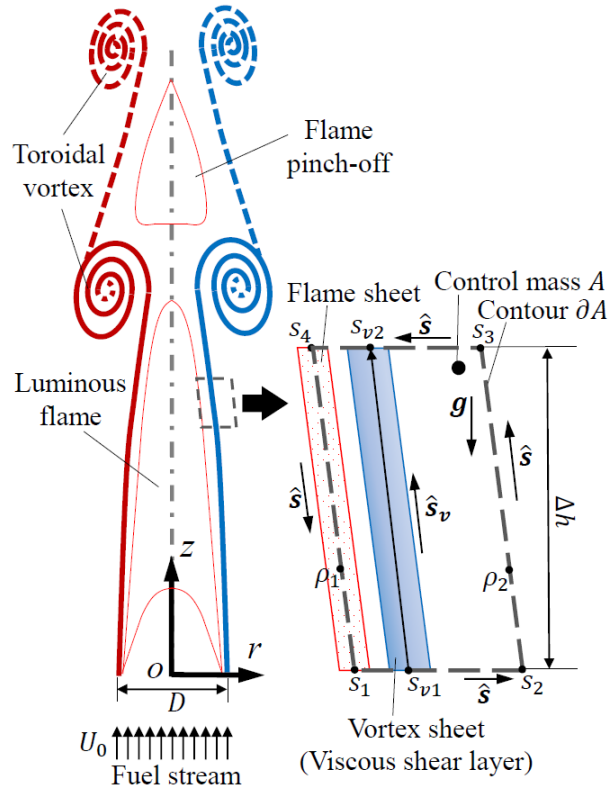

Fig.2 Schematic of the vortex sheet for the growth of the toroidal vortex in a laminar diffusion flame. Note that the zero thickness of flame and vortex sheets are exaggerated for illustration.

A schematic of vortex sheet in a laminar diffusion flame is shown in Fig. 2. To calculate the vorticity growth inside the vortex sheet, we employed the vorticity transport equation in the form of

$$\frac{D\boldsymbol{\omega}}{Dt} = (\boldsymbol{\omega}\cdot\boldsymbol{\nabla})\boldsymbol{u} - \boldsymbol{\omega}(\boldsymbol{\nabla}\cdot\boldsymbol{u}) + \frac{1}{\rho^2}(\boldsymbol{\nabla}\rho\times\boldsymbol{\nabla}p) + \frac{\rho_A}{\rho^2}(\boldsymbol{\nabla}\rho\times\boldsymbol{g}) + \nu\nabla^2\boldsymbol{\omega},\qquad(1)$$

where $\boldsymbol{u}$ and $\boldsymbol{\omega}$ are the velocity and vorticity vectors, $\rho$ and $p$ the local density and gauge pressure, $\boldsymbol{g}$ the gravitational acceleration vector, $\nu$ the kinematic viscosity, and $\rho_A$ the gas density of the ambient environment. It is noted that Ghoniem et al. [31] and Jiang and Luo [10] also used analogs of Eq. (1) in their studies of flame flickering. On the right-hand side of Eq. (1), the first vortex stretching term vanishes for either two-dimensional flows or axisymmetric flows without swirling, the second dilatation term vanishes for incompressible flows, and the fifth diffusion term causes the redistribution of vorticity and thus is not a source of vorticity. Vorticity generation are attributed to the third baroclinic term and the fourth gravitational term, both of which entail the presence of variable density, which is caused by the heat release of combustion. Since the Mach number of the present problem is significantly smaller than unity, the flow field can be treated as isobaric so the third baroclinic term can be neglected. Thus, under the assumptions of laminar axisymmetric incompressible flows without swirling, Eq. (1) is simplified as

$$\frac{D\boldsymbol{\omega}}{Dt} = \frac{\rho_A}{\rho^2}(\nabla \rho \times \boldsymbol{g}) + \nu \nabla^2 \boldsymbol{\omega}, \qquad (2)$$

where the vorticity vector has only one azimuthal component as $\boldsymbol{\omega} = \omega \hat{\boldsymbol{\theta}}$. Eq. (2) can be physically interpreted that the main source of vorticity is a combined effect of density gradient and gravity, which are not parallel to each other.

In studying the formation of a vortex sheet, or physically a viscous shear layer, and its roll-up into a vortex, we are generally interested in the scalar quantity, circulation ($\Gamma$), defined by $\Gamma = \oint_{\partial A} \boldsymbol{u} \cdot d\boldsymbol{s} = \iint \omega dA$. As illustrated in Fig. 2, the dashed box $\partial A$ is a material contour around the vortex sheet segment between $s_{v1}$ and $s_{v2}$, and $A$ is the area encircled by the contour so $A$ is a control mass; $d\boldsymbol{s}$ represents a material line element along $\partial A$. So $\Gamma$ is a measure for the total vorticity inside a vortical structure, and the rate of total change of $\Gamma$ can be derived as

$$\frac{d\Gamma}{dt} = \frac{d}{dt}\oint_{\partial A} \boldsymbol{u} \cdot d\boldsymbol{s} = \oint_{\partial A} \frac{D\boldsymbol{u}}{Dt} \cdot d\boldsymbol{s} + \oint_{\partial A} \boldsymbol{u} \cdot \frac{D(d\boldsymbol{s})}{Dt}. \qquad (3)$$

According to the identity $D(d\boldsymbol{s})/Dt = d\boldsymbol{u}$ given in Ref. [45], the second integral term of Eq. (3) vanishes as $\oint_{\partial A} \boldsymbol{u} \cdot d\boldsymbol{u} = \oint_{\partial A} d(\boldsymbol{u} \cdot \boldsymbol{u})/2 = 0$; the integrand $D\boldsymbol{u}/Dt$ in the first integral term is given by

$$\frac{D\boldsymbol{u}}{Dt} = -\frac{1}{\rho}[\nabla p + (\rho_A - \rho)\boldsymbol{g}] + \nu \nabla^2 \boldsymbol{u}. \qquad (4)$$

To derive Eq. (4) from the Navier-Stokes equation, we have applied the relationship, $\nabla p_{ab} = \nabla p + \rho_A \boldsymbol{g}$, where $p_{ab}$ is the absolute pressure. Note that $\nu \nabla^2 \boldsymbol{u}$ physically represents the shear stress (per unit mass) and approaches zero outside the viscous shear layer, where the flow is effectively inviscid, we have $\nu \nabla^2 \mathbf{u} = \mathbf{0}$ on $\partial A$. Furthermore, under the isobaric approximation, we argue that the pressure gradient is negligible compared with the gravity term for a buoyant flame. With $d\mathbf{s}$ expressed as $\hat{\mathbf{s}} ds$, where $\hat{\mathbf{s}}$ is the unit tangential vector along the contour $\partial A$, Eq. (3) can be derived as

$$\frac{d\Gamma}{dt} = \oint_{\partial A} -\frac{\rho_A - \rho}{\rho} \mathbf{g} \cdot \hat{\mathbf{s}} ds = (\int_{S_2}^{S_3} + \int_{S_4}^{S_1}) \frac{-\rho_A}{\rho} \mathbf{g} \cdot \hat{\mathbf{s}} ds \qquad (5)$$

$$= \int_{S_{v1}}^{S_{v2}} \rho_A \left(\frac{1}{\rho_1} - \frac{1}{\rho_2}\right) (\hat{\mathbf{s}}_v \cdot \mathbf{g}) ds$$

where $\hat{\mathbf{s}}_v$ is the unit tangential vector of the vortex sheet as shown in Fig. 2. In Eq. (5), the line integrals associated with the paths, $s_1$ to $s_2$ and $s_3$ to $s_4$, vanish as $\hat{\mathbf{s}}_v \cdot \mathbf{g} = 0$. It is reasonable to assume that the gas density at either side of the vortex sheet is constant, which means $\rho_1(s) = \rho_1$ and $\rho_2(s) = \rho_2$, and Eq. (5) is further simplified as

$$\frac{d\Gamma}{dt} = \rho_A \left(\frac{1}{\rho_1} - \frac{1}{\rho_2}\right) \left(\int_{S_{v1}}^{S_{v2}} d\hat{\mathbf{s}}_V \cdot \mathbf{g}\right) = -\rho_A g \Delta h \left(\frac{1}{\rho_1} - \frac{1}{\rho_2}\right), \qquad (6)$$

where $\Delta h$ is the vertical distance between the two ends of the vortex sheet element as shown in Fig. 2 Eq. (6) indicates an important feature of a buoyancy-induced vortex sheet that the generation rate of total vorticity (circulation) is irrelevant to the geometric shape of the sheet, but only dictated by the vertical length of the sheet.

Based on Eq. (6), the total circulation of the toroidal vortex grown during one flickering period, $\tau = 1/f$, can be calculated by

$$\Gamma_{TV} = \int_0^\tau \left(\frac{d\Gamma}{dt} + \frac{d\Gamma_0}{dt}\right) dt = \rho_A g H \tau \left(\frac{1}{\rho_2} - \frac{1}{\rho_1}\right) + \int_0^\tau \frac{d\Gamma_0}{dt} dt, \qquad (7)$$

where $\Gamma_{TV}$ is the circulation of the toroidal vortex at the moment of detachment, $\Gamma_0$ the circulation of the initial vortex sheet associated with the fuel jet flow, and $H$ denotes the vertical length of the vortex sheet that is developed during one period and responsible for the roll-up of a single toroidal vortex. The derivation of the second equation assumes that the vortex sheet grows in a quasi-steady manner, given that the flame flickering period is so short that the densities of the gases across the vortex sheet remain approximately unchanged meanwhile. According to previous studies of a starting vortex jet [40], the rate of circulation addition corresponding to the initial jet can be calculated by

$$\frac{d\Gamma_0}{dt} = -C_j U_0^2, \tag{8}$$

where $U_0$ is the initial jet velocity of the fuel at the fire base and $C_j$ is a constant relating to the configuration and conditions of the jet exit. According to the study of Krieg and Mohseni [46], $C_j$ is 0.5 for a parallel-nozzle jet, while its value could be doubled for a converging-nozzle jet with additional circulation contributed from the entrained radial velocity. It is noted that Eq. (8) is applicable to not only a forced jet but also a flow induced by evaporation, for example, a pool fire. During a period of flame flickering, the vortex sheet forming the toroidal vortex is carried by the initial flow to the downstream by $H = U_0 \tau$. Combining Eq. (7) and Eq. (8) yields,

$$\Gamma_{TV} = \int_0^\tau \left(\frac{d\Gamma}{dt} + \frac{d\Gamma_0}{dt}\right) dt = \rho_A g U_0 \tau^2 \left(\frac{1}{\rho_2} - \frac{1}{\rho_1}\right) - C_j U_0^2 \tau. \tag{9}$$

Note that, for the diffusion flame shown in Fig. 2, $\rho_1$ is the density of the gas at the flame sheet, denoted $\rho_f$, whereas $\rho_2$ is the density of the ambient air so that $\rho_2 = \rho_A$. With $\Gamma_{TV}$ scaled by $-U_0 D$, Eq. (9) can be written in the non-dimensional form as

$$\Gamma_{TV}^* = \frac{1}{D}\left[g\tau^2\left(\frac{\rho_A - \rho_f}{\rho_f}\right) + C_j U_0 \tau\right], \tag{10}$$

and further as

$$\Gamma_{TV}^* = Ri\frac{U_0^2 \tau^2}{D^2} + \frac{C_j U_0 \tau}{D} \tag{11}$$

where $Ri$ is the Richardson number defined as $Ri = (1 - \rho_A/\rho_f)gDU_0^{-2}$.

At this point, we recall that the flame flickering is caused by the alternate formation and detachment of the toroidal vortex. The analysis hitherto has addressed the formation and growth of the toroidal vortex, which leaves us the other half of the question: how does the toroidal vortex shed? We noted that similar problems have been considered by previous studies on vortex ring formation [43, 44]. These studies point to the existence of a universal non-dimensional formation number (a typical value of 4 for an ideal starting vortex jet, but could change notably under various conditions [47-50]), above which a vortex ring would be too strong to maintain growing and consequently it detaches from its vorticity-feeding shear layer. In this sense, the formation number can be considered as a dimensionless measure of the upper limit of the total circulation of an attached vortex ring.

Based on the above considerations, we hypothesize that the shedding of the toroidal vortex is controlled by the same mechanism: the continuous growing of the shear layer causes the gradual accumulation of the circulation inside the toroidal vortex until it reaches a threshold denoted by $C$. Applying $\Gamma_{TV}^* = C$ to Eq. (11) yields

$$f = \frac{1}{\tau} = \frac{1}{2C}\sqrt{\frac{(\rho_A - \rho_f)g}{\rho_f D}}\left(\sqrt{\frac{C_j^2}{Ri}} + \sqrt{\frac{C_j^2}{Ri} + 4C}\right), \tag{12}$$

which completes the derivation for the frequency of a general buoyancy-driven diffusion flame. It is noted that Eq. (12) has almost the same form as the scaling formula given by Cetegen and Ahmed (Eq. 1 of Ref. [21]), except that Eq. (12) includes two additional coefficients, $C_j$ related to the formation of a toroidal vortex and $C$ to the detachment of the vortex.

For a pool fire where fuel vapor enters the system through the evaporation of liquid fuel, the initial fuel velocity $U_0$ may be considered to be negligible compared with the characteristic velocity ($\sqrt{gD}$) induced by gravity, rendering $Ri \to \infty$. Consequently, Eq. (12) can be simplified as

$$f_p = \sqrt{\frac{(\rho_A - \rho_f)g}{C\rho_f D}} = \sqrt{\frac{(r^* - 1)g}{CD}}, \tag{13}$$

where $r^* = \rho_A/\rho_f$. Again, the functionality of Eq. (13) is consistent with the scaling law obtained by Byram and Nelson [15], verifying that the pool fire can be approximately treated as a limiting case of the general diffusion flame. In the following section, we shall validate Eq. (12) and Eq. (13) and discuss the choices of the coefficients, $C_j$, $C$, and $r^*$, through the comparisons with previous experimental and numerical data of buoyant pool and jet flames.

**III. Determination of Threshold Value for Vortex Detachment**

In this section, we first predict the flickering frequency of pool fires given by Eq. (13). As discussed in the Introduction, the scaling law, $f_p \propto D^{-1/2}$, has been demonstrated by numerous previous studies, but the relationship $f_p \propto (g/D)^{1/2}$ has not been sufficiently verified, especially for the cases with varying gravity. We plotted $f_p$ against $\sqrt{g/D}$ in Fig. 3 for various pool fires from existing literature, which have different fuel types, fire source dimensions, and gravities (the variation ranges are displayed in the legend of Fig. 3). It is noted that the original experiment of Sato et al. [51] was performed mainly for jet flames. Here, we adopted their low-velocity jet flames as approximations for pool flames. This may be justified by the findings of Durox et al. [52] and Sato et al. [51] that the frequency is relatively independent of fuel velocity for a low-velocity jet flame, where the Froude number is low and buoyancy dominates over the initial jet flow.

In general, Fig. 3 confirms that the existing experimental data together the current result are in good agreement with the trend line, $f_p = 0.48\sqrt{g/D}$ with $g$ being 9.8 m/s², which is equivalent to the scaling relation, $f = 1.5\sqrt{1/D}$, given by Cetegen and Ahmed [21]. Based on this trend line, the coefficients in Eq. (13) are correlated to each other as $C = 4.34(r^* - 1)$. To estimate $r^*$, we assume the thermochemical process at the flame sheet to be isobaric. Then, following the ideal gas law, the densities are related to the temperatures as $\rho_A/\rho_f = T_f/T_A = r^*$, where $T_A$ and $T_f$ are the temperatures of the ambient air and the flame sheet, respectively. For the different types of fuel involved in Fig. 3, the flame temperature at normal atmospheric condition (1 atm, $T_A \approx 300$ K) varies in a wide range around 2000 K, corresponding to $r^* \approx 6.7$ and $C \approx 25$. As discussed previously, $C$ physically represents the threshold value of the non-dimensional circulation held up by a rolled-up vortex, so the current relationship of $C$ obtained based on pool flames should be applicable to the formation of toroidal vortex in a general diffusion flame, including the jet flames which will be analyzed in the next section. Here, it should be clarified that although $C$ shares the same physical essence as the formation number of a starting vortex jet, their values are not directly comparable to each other due to the different characteristic scales used in the nondimensionalization.

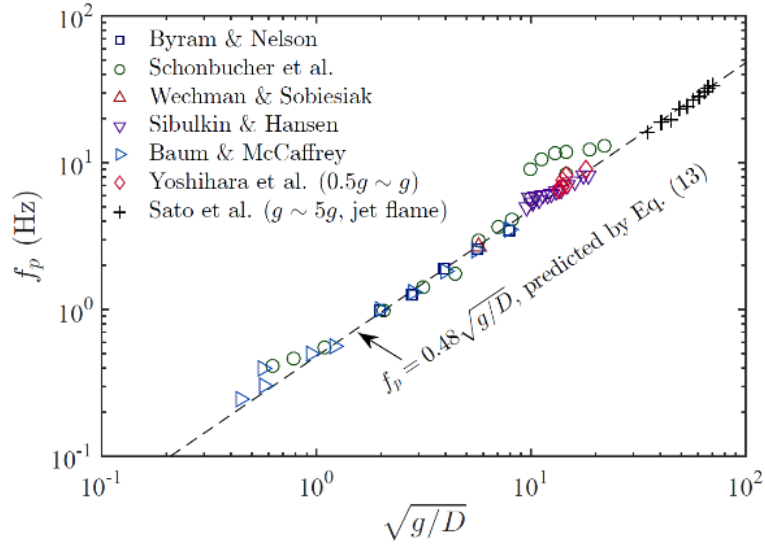

Fig. 3 Flickering frequencies of pool flames ($f_p$) as a function of $\sqrt{g/D}$, with experimental data taken from Refs. [17, 51, 53-56]. Note that Sato et al.'s low-speed jet flames [51] were adopted as approximations of pool flames.

**IV. Model Predictions for Jet Flames**

For jet diffusion flames, the effect of fuel velocity on the flickering frequency has been investigated by Harmins et al. [22], Sato et al. [51], and Fang et al. [57], among others. Cetegen and Ahmed [21] proposed a scaling formula similar to Eq. (12) of the current study and experimentally verified its functionality for jet flames and isothermal plumes. However, to the authors' knowledge, their formula of $f$ vs. $Ri$ has not been substantiated by later studies of jet flames. A possible reason might be that the calculation of the Richardson number requires calculating the density ratio $r^*$, which is somewhat arbitrary in practice, and any uncertainty could easily cause the deviation of the entire data set. To eliminate the ambiguity of evaluation of $Ri$, we rewrote Eq. (12) in the form of

$$St = \frac{1}{2C}\left(C_j + \sqrt{C_j^2 + 4C(r^* - 1)/Fr}\right), \tag{14}$$

where the Strouhal number and the Froude number are defined as $St = fD/U_0$ and $Fr = U_0^2/gD$, respectively. Because $r^*$ is not contained in the definitions of $St$ or $Fr$, they can be estimated with sufficient accuracy for jet flame experiments. Here, $C_j = 0.5$ for a canonical starting vortex jet [40] is applied. However, it is noted that $C_j$ for a jet flame could be different from a starting vortex jet, considering that the thermal expansion at the exit of the jet flame could induce a radial velocity which would contribute to additional circulation. From Section **III**, the relationship $C = 4.34(r^* - 1)$ obtained from pool flames should also hold for jet flames and is adopted here as a known condition, so $r^*$ is the only unknown parameter in Eq. (14). Based on the previous discussion, $r^*$ is a variable around 6.7, and is set to be 4, 7, and 10, respectively, to evaluate its effect on this model.

Now, the relation of $St$ vs. $1/Fr$ is plotted in Fig. 4 for the jet diffusion flames from several previous studies, displaying generally promising agreement with the models given by Eq. (14), especially in the $1/Fr > 1$ region where the jet is dominated by buoyancy. It is interesting to note that the influence of $r^*$ disappears in the $1/Fr > 1$ region as the different models converge to a single line. This again verifies that in the limit $Fr \to 0$ the effect of the initial jet becomes negligible and the solution approaches that of a pool flame.

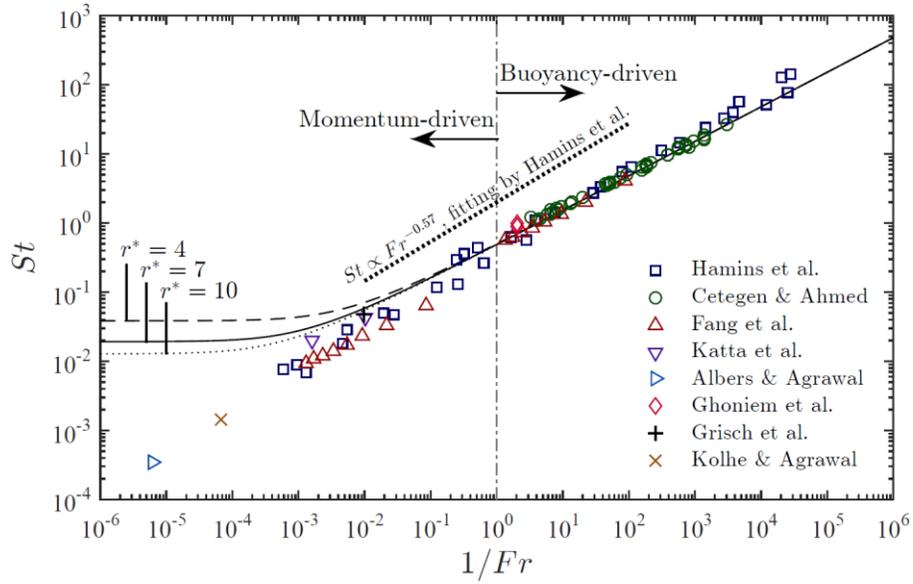

Fig. 4 The Strouhal number as a function of $1/Fr$ for different jet diffusion flames, with experimental data taken from Refs. [21, 22, 30-33, 57, 58]. The current model predictions of Eq. (14) with $r^*$ being 4, 7, and 10, are presented. The scaling law, $St \propto Fr^{-0.57}$, reported by Harmins et al. [22] is also plotted for comparison.

For comparison, we also plot the scaling relation of Harmins et al. [22], which was obtained based on fitting the entire data span. It is seen that their fitting line, with an exponent of $-0.57$, seems to capture the trend slope better in the $1/Fr < 1$ region than the $1/Fr > 1$ region. Actually, our model shows that in the $1/Fr > 1$ region the exponent of the scaling law should approximate exactly $-0.5$. This indicates that the flickering of diffusion jet flames might have two different regimes, momentum-driven and buoyancy-driven, which are also evident from the studies of Sato et al. [51] and Fang et al. [57]. Regarding the momentum-driven region, Eq. (14) predicts that $St$ asymptotically approaches the constant $C_j/C$ as $Fr \to \infty$. As shown in Fig. 4, the mismatch between the current model and experimental data at large $Fr$ indicates the existence of a different mechanism for momentum-driven jet flames, which raises a question for future work.

## V. Conclusions

The classic problem of flickering laminar diffusion flames was theoretically revisited in this study by performing vortex dynamics analysis. Considering the previous experimental observations that the flame flickering is synchronized with the periodic toroidal vortices, we sought to provide further mathematical modeling for the growth and detachment of the toroidal vortex, both of which are essential contributions to the periodicity of the flow field. By calculating the total circulation of a toroidal vortex grown during one period and equating that to a non-dimensional universal constant responsible for the vortex detachment, we analytically derived a general formula for the flickering frequency of buoyant diffusion flames. This formula not only confirms the previous scaling relation of Cetegen and Ahmed [21] but also rigorously correlate the dynamics of the toroidal vortex and flame flickering, therefore presenting the complete physical picture of the problem. This is reflected by the physically definite parameters $C_j$ and $C$ that account for the vortex growth from the initial jet and the detachment of the toroidal vortex, respectively. Expressed in terms of the Strouhal and Froude numbers to eliminate the arbitrariness of Richardson number, the formula shows convincing agreement with experimental and numerical data from existing literature, especially for pool flames and jet diffusion flames with low Froude number. The future development of the current theory of buoyant diffusion flames can involve other mechanisms to capture the frequencies of jet diffusion flames with large Froude number, when gravitational effect becomes secondary.

**Acknowledgment**

This work was supported partly by National Natural Science Foundation of China (Grant No. 91641105) and partly by the Hong Kong Polytechnic University (G-UA2M and G-YBGA). The authors are grateful for an additional support by a collaborative open fund from the Key Laboratory of High-temperature Gas Dynamics, Chinese Academy of Sciences.

**References**

[1] S.P. Burke, T.E.W. Schumann, Industrial & Engineering Chemistry, 20 (1928) 998-1004.
[2] A. Linan, M. Vera, A.L. Sanchez, Annual Review of Fluid Mechanics, 47 (2015) 293-314.
[3] L.D. Chen, J.P. Seaba, W.M. Roquemore, L.P. Goss, Proceedings of the Combustion Institute, 22 (1989) 677-684.
[4] G. Cox, Combustion fundamentals of fire, Academic Press, 1995.
[5] C.M. Coats, Progress in Energy and Combustion Science, 22 (1996) 427-509.
[6] A. Lingens, K. Neemann, J. Meyer, M. Schreiber, Proceedings of the Combustion Institute, 26 (1996) 1053-1061.
[7] S.R. Tieszen, V.F. Nicolette, L.A. Gritzo, J. Moya, J. Holen, D. Murray, in, Sandia National Labs., Albuquerque, NM, 1996.
[8] W.M.G. Malalasekera, H.K. Versteeg, K. Gilchrist, Fire and Materials, 20 (1996) 261-271.
[9] P. Joulain, Proceedings of the Combustion Institute, 27 (1998) 2691-2706.
[10] X. Jiang, K.H. Luo, Proceedings of the Combustion Institute, 28 (2000) 1989-1995.
[11] S.R. Tieszen, Annual review of fluid mechanics, 33 (2001) 67-92.
[12] A. Linan, E. Fernandez-Tarrazo, M. Vera, A.L. Sanchez, Combustion Science and Technology, 177 (2005) 933-953.
[13] D.S. Chamberlin, A. Rose, Proceedings of the Combustion Institute, 1 (1948) 27-32.
[14] C.S. McCamy, Journal of Research of the National Bureau of Standards, 56 (1956) 293.
[15] G.M. Byram, R.M. Nelson, Fire Technology, 6 (1970) 102-110.
[16] R. Portscht, Combustion Science and Technology, 10 (1975) 73-84.
[17] M. Sibulkin, A.G. Hansen, Combustion Science and Technology, 10 (1975) 85-92.
[18] P.h. Detriche, J.C. Lanore, Fire Technology, 16 (1980) 204-211.
[19] E.E. Zukoski, B.M. Cetegen, T. Kubota, Proceedings of Combustion Institute, 20 (1985) 361-366.
[20] L.M. Trefethen, R.L. Panton, Applied Mechanics Reviews, 43 (1990) 153-170.
[21] B.M. Cetegen, T.A. Ahmed, Combustion and Flame, 93 (1993) 157-184.
[22] A. Hamins, J.C. Yang, T. Kashiwagi, Proceedings of the Combustion Institute, 24 (1992) 1695-1702.
[23] A. Bejan, Journal of Heat Transfer (Transcations of the ASME, Series C), 113 (1991) 261-263.
[24] J. Buckmaster, N. Peters, Proceedings of the Combustion Institute, 21 (1988) 1829-1836.
[25] L.D. Chen, W.M. Roquemore, Combustion and Flame, 66 (1986) 81-86.
[26] W.M. Roquemore, L.D. Chen, J.P. Seaba, P.S. Tschen, L.P. Goss, D.D. Trump, Physics of Fluids (1958-1988), 30 (1987) 2600.
[27] R.W. Davis, E.F. Moore, R.J. Santoro, J.R. Ness, Combustion Science and Technology, 73 (1990) 625-635.
[28] J. Carpio, M. Sánchez-Sanz, E. Fernández-Tarrazo, Combustion and Flame, 159 (2012) 161-169.
[29] V.R. Katta, W.M. Roquemore, Combustion and Flame, 92 (1993) 274-282.
[30] V.R. Katta, L.P. Goss, W.M. Roquemore, AIAA journal, 32 (1994) 84-94.
[31] A.F. Ghoniem, I. Lakkis, M. Soteriou, Proceedings of the Combustion Institute, 26 (1996) 1531-1539.
[32] B.W. Albers, A.K. Agrawal, Combustion and flame, 119 (1999) 84-94.
[33] P.S. Kolhe, A.K. Agrawal, Flow, Turbulence and Combustion, 79 (2007) 343-360.
[34] W.E. Mell, K.B. McGrattan, H.R. Baum, Proceedings of the Combustion Institute, 26 (1996) 1523-1530.
[35] B.M. Cetegen, Physics of Fluids, 9 (1997) 3742-3752.
[36] T. Maxworthy, Journal of Fluid Mechanics, 390 (1999) 297-323.
[37] T. Maxworthy, Journal of Fluid Mechanics, 51 (1972) 15-32.
[38] T. Maxworthy, Journal of Fluid Mechanics, 81 (1977) 465-495.

[39] P. Saffman, Journal of Fluid Mechanics, 84 (1978) 625-639.
[40] N. Didden, Zeitschrift für Angewandte Mathematik und Physik (ZAMP), 30 (1979) 101-116.
[41] A. Glezer, The Physics of fluids, 31 (1988) 3532-3542.
[42] K. Shariff, A. Leonard, Annual Review of Fluid Mechanics, 24 (1992) 235-279.
[43] M. Gharib, E. Rambod, K. Shariff, J. Fluid Mech., 360 (1998) 121-140.
[44] K. Mohseni, M. Gharib, Physics of Fluids, 10 (1998) 2436-2438.
[45] J.-Z. Wu, H.-Y. Ma, M.-D. Zhou, Vorticity and vortex dynamics, Springer Science & Business Media, 2007.
[46] M. Krieg, K. Mohseni, Journal of Fluid Mechanics, 719 (2013) 488-526.
[47] J.O. Dabiri, M. Gharib, Journal of Fluid Mechanics, 538 (2005) 111-136.
[48] P.S. Krueger, J.O. Dabiri, M. Gharib, Journal of Fluid Mechanics, 556 (2006) 147-166.
[49] J.M. Lawson, J.R. Dawson, Physics of Fluids, 25 (2013) 105113.
[50] X. Xia, K. Mohseni, Physics of Fluids, 27 (2015) 115101.
[51] H. Sato, K. Amagai, M. Arai, Combustion and Flame, 123 (2000) 107-118.
[52] D. Durox, T. Yuan, F. Baillot, J.M. Most, Combustion and flame, 102 (1995) 501-511.
[53] H.R. Baum, B.J. McCaffrey, Fire Safety Science, 2 (1989) 129-148.
[54] A. Schönbucher, B. Arnold, V. Banhardt, V. Bieller, H. Kasper, M. Kaufmann, R. Lucas, N. Schiess, in: Symposium (International) on Combustion, Elsevier, 1988, pp. 83-92.
[55] E.J. Weckman, A. Sobiesiak, in: Symposium (International) on Combustion, Elsevier, 1989, pp. 1299-1310.
[56] N. Yoshihara, A. Ito, H. Torikai, Proceedings of the Combustion Institute, 34 (2013) 2599-2606.
[57] J. Fang, J.-w. Wang, J.-f. Guan, Y.-m. Zhang, J.-j. Wang, Fuel, 163 (2016) 295-303.
[58] F. Grisch, B. Attal-Tretout, P. Bouchardy, V.R. Katta, W.M. Roquemore, Journal of Nonlinear Optical Physics & Materials, 5 (1996) 505-526.